\documentclass[preprints,article,accept,moreauthors,pdftex,10pt,a4paper]{Definitions/mdpi}
\usepackage{subfigure}

\firstpage{1} 
\pubvolume{xx}
\issuenum{1}
\articlenumber{5}
\pubyear{2018}
\copyrightyear{2018}
\history{}

\pdfoutput=1

\Title{Testing a best-fit hydrodynamical model using PCA$^{\dagger}$}

\Author{Tiago Nunes da Silva $^{1*}$\orcidA{}, David Dobrigkeit Chinellato$^{1}$\orcidB{}, Rafael Derradi de Souza$^{1}$\orcidD{}, Maur\'{i}cio Hippert $^{2}$,  Matthew Luzum$^{2}$\orcidC{},  Jorge Noronha$^{2}$  and Jun Takahashi $^{1}$}

\AuthorNames{Tiago Nunes da Silva, David Dobrigkeit Chinellato, Rafael Derradi de Souza, Mauricio Hippert, Matthew Luzum, Jorge Noronha and Jun Takahashi}

\address{%
$^{1}$ \quad Universidade Estadual de Campinas, S\~{a}o Paulo, Brazil;\\
$^{2}$ \quad Universidade de S\~{a}o Paulo, S\~{a}o Paulo, Brazil;}

\corres{Correspondence: tiagoj.nunes@gmail.com;}

\firstnote{Presented at Hot Quarks 2018 - Workshop for young scientists on the physics of ultrarelativistic nucleus-nucleus collisions, Texel, The Netherlands, September 7-14 2018}

\abstract{Recently, a comprehensive Bayesian analysis was performed to simultaneously extract the values of a number of hydrodynamic parameters necessary for compatibility with a limited set of experimental data from the LHC. In this work, this best-fit model is tested against newly measured experimental flow results not included in the original work, namely the  principal components of the two-particle correlation matrix in transverse momentum. The results from simulations show a good numerical agreement with data obtained by the CMS Collaboration.}

\keyword{Heavy-ion collisions; quark-gluon plasma; relativistic fluid dynamics; principal component analysis.}

\begin{document}

\section{Introduction}

Relativistic heavy-ion collision (HIC) experiments have proven to be an important tool in exploring the fundamental nature of strongly interacting matter under extreme conditions. The standard picture of the processes involved in such collisions is that after a short period of time following the collision of the original nuclei, the evolution of the resulting system can be described by relativistic viscous hydrodynamics. In fact, simulations of heavy-ion collisions based on hydrodynamical evolution are able to describe several observables from experimental data with great accuracy \cite{Heinz:2013th}. The hydrodynamical evolution can be complemented by a subsequent simulation of hadronic cascade models in order to describe the evolution of the gas of hadrons formed after the fluid cools down and the particles hadronize. This combination is usually referred to as a \textit{hybrid model} \cite{Petersen:2008dd}.

Recently, Bernhard \textit{et al}. used a hybrid model \cite{Bernhard:2016tnd, Bernhard:2018hnz} consisting of the TRENTo model \cite{Moreland:2014oya} for the generation of initial conditions, the VISH2+1 code for hydrodynamical evolution, and the UrQMD transport model \cite{Bass:1998ca, Bleicher:1999xi} for the evolution of the hadron gas phase. Through a Bayesian analysis the authors have obtained the optimal a posteriori values for a series of parameters required by the model. In this work we utilize a similar setup and test the validity of the model with these parameters for a set of new observables not included in the original Bayesian analysis, namely results from a principal component analysis (PCA) of the two-particle correlation matrix in transverse momentum \cite{Bhalerao:2014mua}. 

\section{Results}

We have performed event by event simulations of collisions between Pb nuclei at energies $\sqrt{s_{NN}}=2.76$ TeV and $\sqrt{s_{NN}}=5.02$ TeV. A sample of one million initial conditions was generated for centrality calibration, which was based on total entropy. Because of the strong correlation between entropy and final charged particle multiplicity, this definition is essentially equivalent to what is done experimentally. The resulting entropy distribution allows us to classify the centrality of a given event.

\begin{figure}
\centering
\includegraphics[width=0.8\columnwidth]{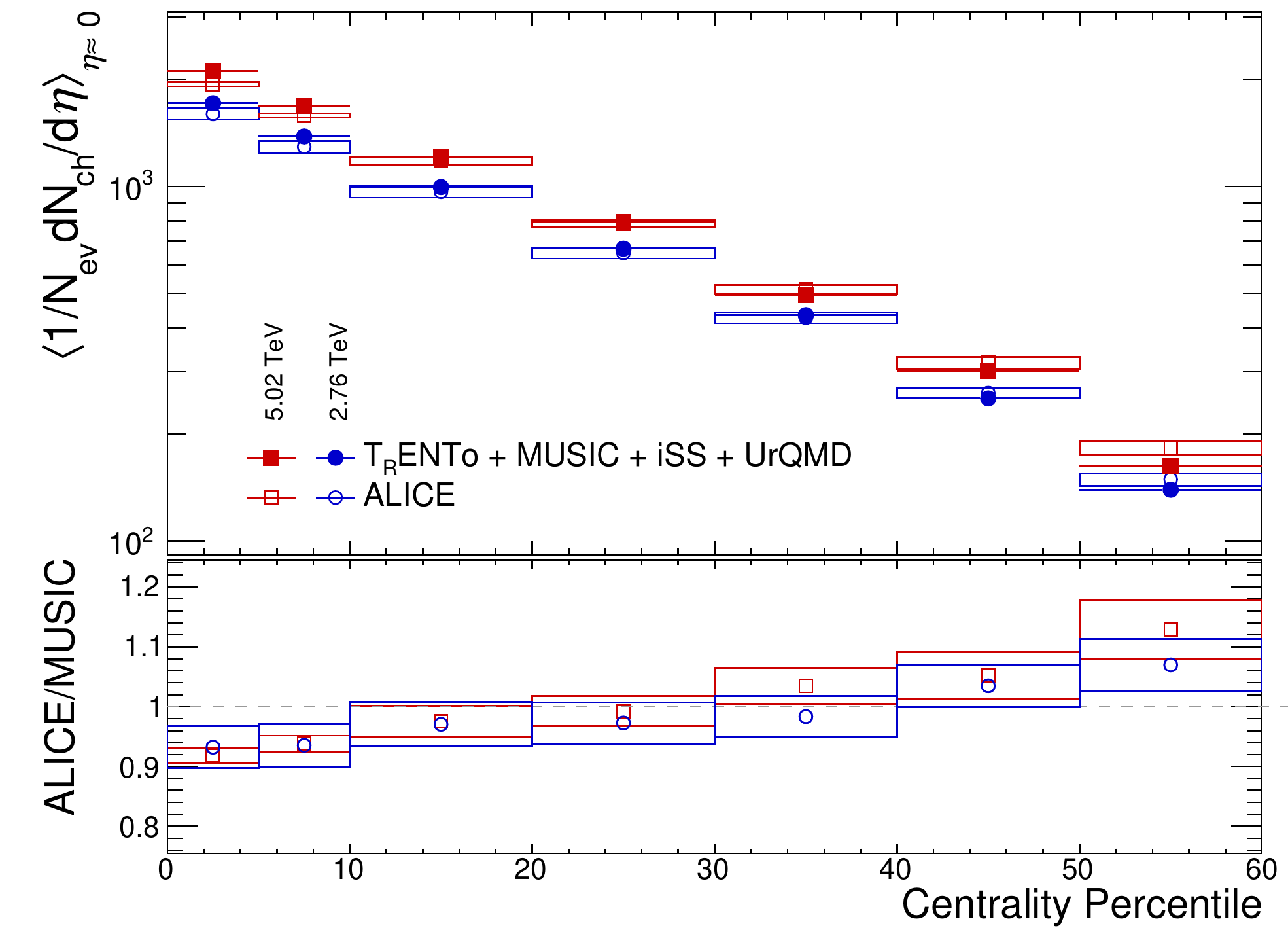}
\caption{Charged particle multiplicity from simulations of Pb-Pb collisions at $\sqrt{s_{NN}}=2.76$ TeV and $\sqrt{s_{NN}}=5.02 $ TeV as a function of centrality. Results are compared to data from the ALICE Collaboration\cite{Aamodt:2010cz, Adam:2015ptt}.}
\label{fig:multi-vs-cent}
\end{figure}

The resulting charged particle multiplicity distribution as a function of event centrality is presented in Figure~\ref{fig:multi-vs-cent}. The results from simulations are compared to data from the ALICE Collaboration \cite{Aamodt:2010cz, Adam:2015ptt}. Up to the centrality bin corresponding to 50\% to 60\%, simulation results agree with experimental data to 10\% accuracy.

\begin{figure}[H]
  \centering
  \subfigure[]{\includegraphics[width=0.49\columnwidth]{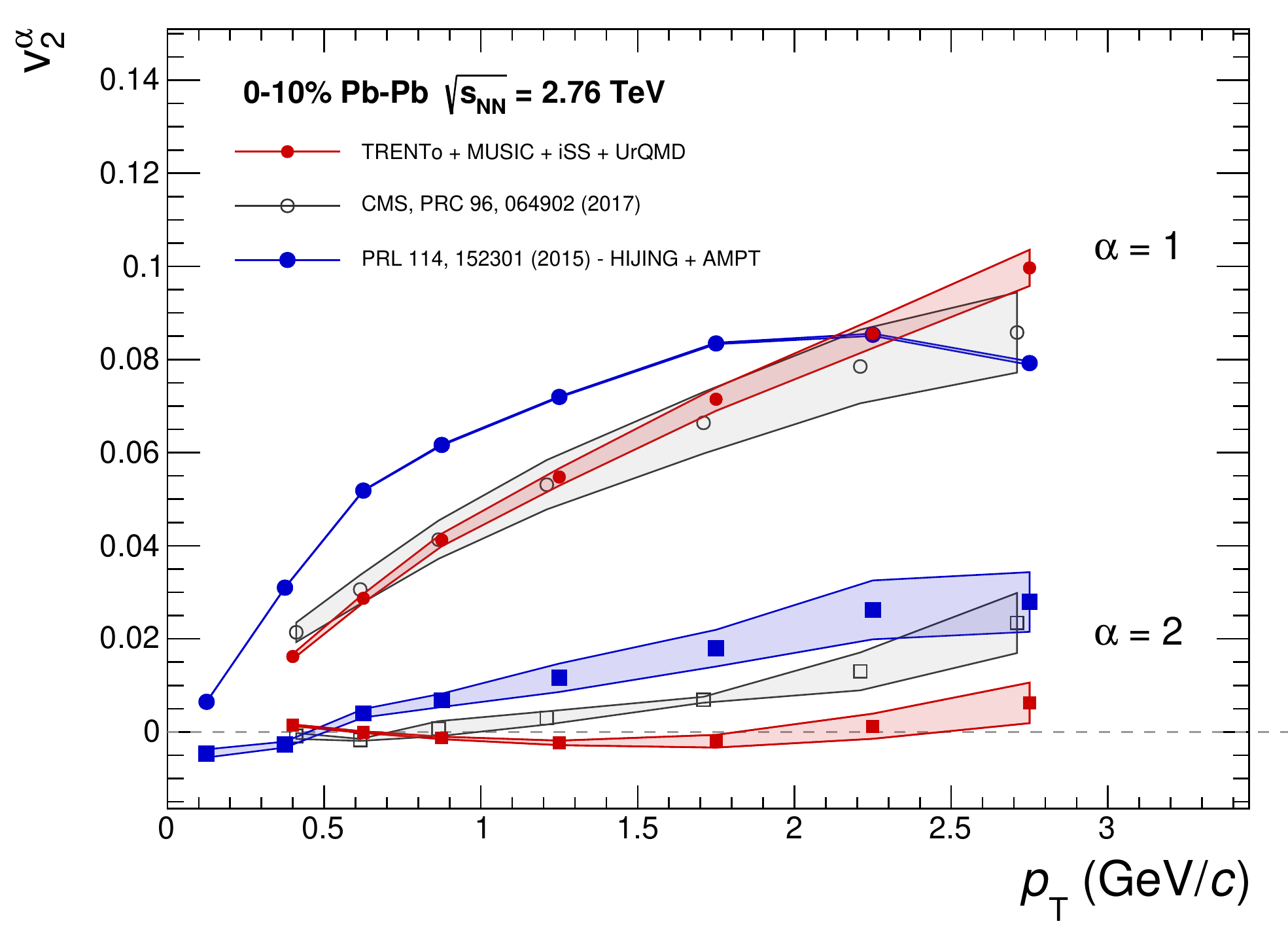}\label{fig:pca-v2}}
  \subfigure[]{\includegraphics[width=0.49\columnwidth]{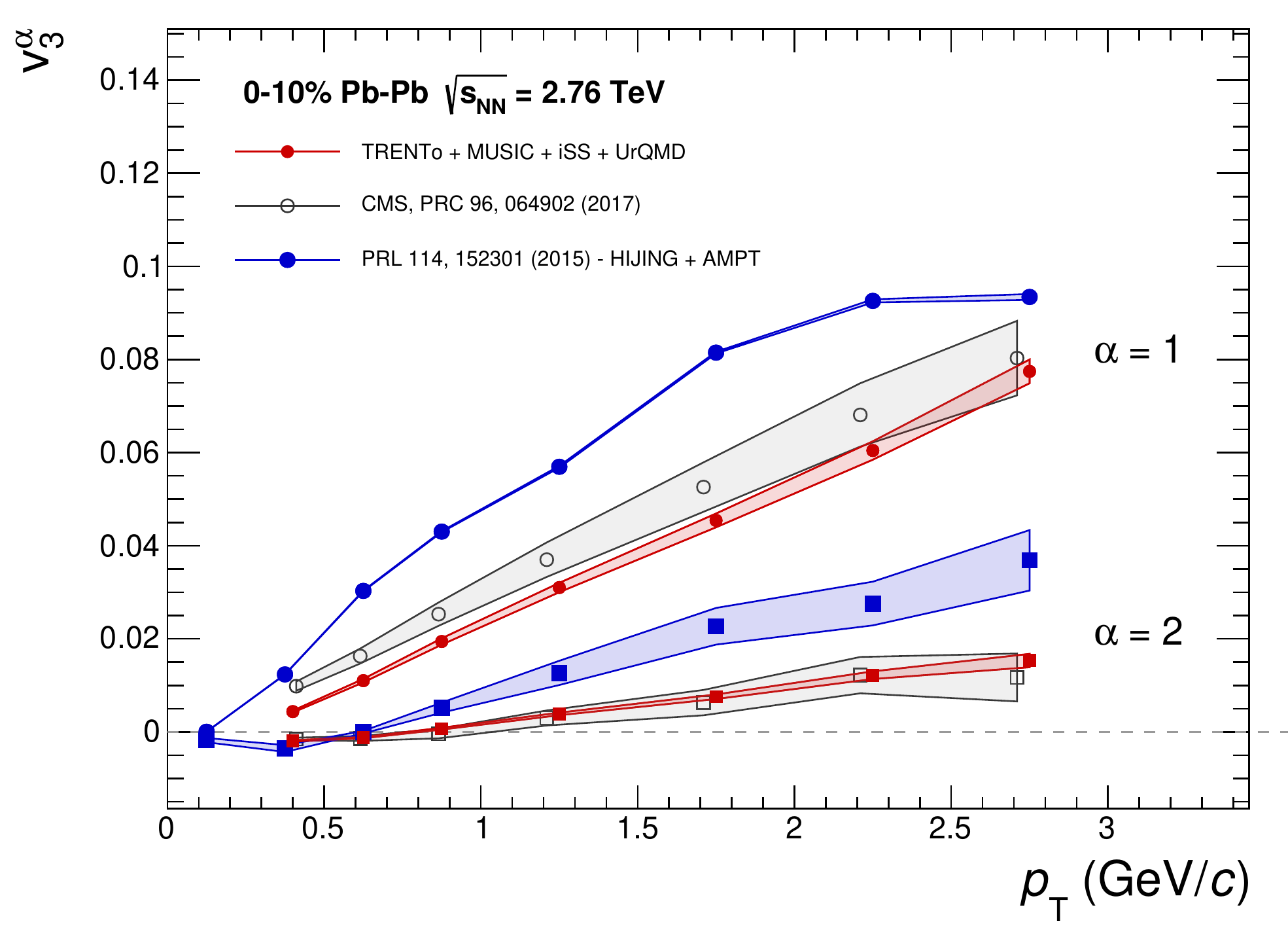}\label{fig:pca-v3}}
  \caption{Principal components of the two-particle correlation matrix for Pb-Pb collisions at center of mass energy $\sqrt{s_{NN}} = 2.76$ TeV and 0\% to 10\% centrality. Results from this work (red) are compared to the original PCA work \cite{Bhalerao:2014mua} (blue) and to data from the CMS Collaboration \cite{Sirunyan:2017gyb} (gray).}
  \label{fig:pca}
\end{figure}

The main result of this work is the first calculation of PCA of the two-particle correlation matrix in transverse momentum using a hybrid model and realistic hydrodynamical initial conditions. A subset of these results is shown in Figure~\ref{fig:pca} (full results will be reported in a forthcoming work). 

\section{Materials and Methods}

The hybrid model for HIC simulations devised in this work is comprised of the following ingredients:

\begin{itemize}
    \item TRENTo, a parametric wounded nucleon model for generation of initial conditions for hydrodynamics in HIC \cite{Moreland:2014oya};
    \item MUSIC, an Eulerian 3D+1 relativistic second-order viscous hydrodynamics code for event by event HIC simulations \cite{Schenke:2010nt, Schenke:2011bn};
    \item UrQMD, a transport model for the evolution of the hadron gas \cite{Bass:1998ca, Bleicher:1999xi};
\end{itemize}

Parameter values for the simulation chain were obtained from the work by Bernhard \textit{et al.}~\cite{Bernhard:2018hnz}. We have chosen not to include in our simulation chain a period of free streaming between the initial condition generation and the beginning of the hydrodynamical evolution, i.e., we utilize the output from TRENTo as the initial distribution of entropy density for hydrodynamics. Because of that, the overall normalization constant for TRENTo had to be recalculated, which was done by matching the charged-particle multiplicity density to experimental data from the ALICE Collaboration \cite{Aamodt:2010cz, Adam:2015ptt}. 
We have also developed a ROOT-based C++ class for storing the resulting final configuration of particles from each event, called HadrEx. 

The Q-cumulants of the event sample are built in bins of transverse momentum following the definition by Bhalerao \textit{et al.} \cite{Bhalerao:2014mua}
\begin{equation}
    Q_n (p) \equiv \frac{1}{2\pi\,\Delta p_t \,\Delta \eta}\sum_{j=1}^{M(p)}\exp{(i n \varphi_j)}.
\end{equation}
The pair distribution is then obtained through the formula
\begin{equation}
    V_{n\Delta}(p_a, p_b) \equiv \langle Q_n(p_a) Q_n^* (p_b)\rangle - \frac{\langle M(p_a) \rangle \delta_{p_a, p_b}}{(2\pi \, \Delta p_t \, \Delta\eta)^2}, 
\end{equation}
where the first term is the two-particle correlation matrix and the second term removes self-correlations.

The principal components are calculated by diagonalizing the matrix $V_{n\Delta}(p_a, p_b)$ and identifying the results with the PCA approximation
\begin{equation}
    V_{n \Delta} (p_a, p_b) = \sum_\alpha \lambda^{(\alpha)}\psi^{(\alpha)}(p_a)\psi^{(\alpha)*}(p_b) \approx \sum_{\alpha = 1}^k V_n^{(\alpha)}(p_a) V_n^{(\alpha)*}(p_b),
\end{equation}
so that
\begin{equation}
    V_n^{(\alpha)} (p) \equiv \sqrt{\lambda^\alpha} \psi^{(\alpha)}(p) \,\,\,\, \text{and} \,\,\,\, v_n^{(\alpha)}(p) \equiv \frac{ V_n^{(\alpha)} (p)}{\langle V_0(p)\rangle}
\end{equation}
express the principal component $\alpha$ of the $n$-th harmonic of the anisotropic flow in terms of the eigenvalue $\alpha$ and its associated eigenvector, with the eigenvalues ordered from largest to smallest absolute value. The normalization in the second equation allows for a direct comparison with the usual measurement of the differential flow.


\section{Discussion}

The principal components of the two-particle correlation matrix are an interesting observable to study: by considering the full covariance matrix, including the off-diagonal terms correlating particles lying in different bins of transverse momentum,  they contain more information about the anisotropic flow than the usual measurements of the differential flow coefficients via two-particle correlations, which only correlate particles in the same transverse momentum bin. The $p_T$-dependent event-by-event fluctuations break factorization of the pair distribution
\begin{equation}
    \left\langle \frac{dN_{pairs}}{\vec{dp_a} \vec{dp_b}} \right\rangle = \left\langle \frac{dN}{\vec{dp_a}} \frac{dN}{\vec{dp_b}} \right\rangle
\end{equation}
into a product of single-particle probability distributions (as expected in pure hydrodynamics simulations without event-by-event fluctuations). The subleading components of the PCA measure the size of these fluctuations \cite{Bhalerao:2014mua}.

 It is noteworthy that, even though these observables were not included in the original Bayesian analysis, the results from the simulations are in good agreement with experimental results from the CMS Collaboration \cite{Sirunyan:2017gyb}, as observed in Figure~\ref{fig:pca}, extending the validity of the hybrid model. In future work we will also present an extended analysis of further observables (such as symmetric cumulants) and extend this simulation framework to consider lower energies and small systems.

\vspace{6pt} 

\authorcontributions{Conceptualization, all; methodology, T.N.dS., M.H. and M.L.; software, T.N.dS., D.C., R.D.dS. and M.L.; validation, T.N.dS., D.C., R.D.dS. and M.H.; formal analysis, T.N.dS. and D.C.; investigation, all; resources, D.C., M. L. and J.T.; data curation, T.N.dS. and M.L.; writing—original draft preparation, T.N.dS.; writing—review and editing, all; visualization, T.N.dS and D.C; supervision, D.C., M.L., J.N. and J.T.; project administration, J.T.; funding acquisition, J.T.}

\funding{This research was funded by FAPESP grants number  2014/09167-8 (R.D.dS.), 2016/13803-2 (D.D.C.), 2016/24029-6 (M.L.), 2017/05685-2 (all), 2018/01245-0 (T.N.dS.) and 2018/07833-1(M.H.). D.D.C., M.L., J.N., and J.T. thank CNPq for financial support.}

\acknowledgments{We would like to acknowledge computing time provided
by the Research Computing Support Group at Rice University through agreement with the University of S\~{a}o Paulo.}

\conflictsofinterest{The authors declare no conflict of interest.} 

\externalbibliography{yes}
\bibliography{ref.bib}

\begin{thebibliography}{-------}
\providecommand{\natexlab}[1]{#1}

\bibitem[Heinz and Snellings(2013)]{Heinz:2013th}
Heinz, U.; Snellings, R.
\newblock {\em Ann. Rev. Nucl. Part. Sci.} {\bf 2013}, {\em 63},~123--151.

\bibitem[Petersen \em{et~al.}(2008)Petersen, Steinheimer, Burau, Bleicher, and
  Stocker]{Petersen:2008dd}
Petersen, H.; Steinheimer, J.; Burau, G.; Bleicher, M.; Stocker, H.
\newblock {\em Phys. Rev.} {\bf 2008}, {\em C78},~044901.

\bibitem[Bernhard \em{et~al.}(2016)Bernhard, Moreland, Bass, Liu, and
  Heinz]{Bernhard:2016tnd}
Bernhard, J.E.; Moreland, J.S.; Bass, S.A.; Liu, J.; Heinz, U.
\newblock {\em Phys. Rev.} {\bf 2016}, {\em C94},~024907.

\bibitem[Bernhard(2018-04-19)]{Bernhard:2018hnz}
Bernhard, J.E.
\newblock PhD thesis, Duke U.,  2018-04-19,
  \href{http://xxx.lanl.gov/abs/1804.06469}{{\normalfont
  [arXiv:nucl-th/1804.06469]}}.

\bibitem[Moreland \em{et~al.}(2015)Moreland, Bernhard, and
  Bass]{Moreland:2014oya}
Moreland, J.S.; Bernhard, J.E.; Bass, S.A.
\newblock {\em Phys. Rev.} {\bf 2015}, {\em C92},~011901.

\bibitem[Bass \em{et~al.}(1998)Bass et~al.]{Bass:1998ca}
Bass, S.A.; others.
\newblock {\em Prog. Part. Nucl. Phys.} {\bf 1998}, {\em 41},~255--369.

\bibitem[Bleicher \em{et~al.}(1999)Bleicher et~al.]{Bleicher:1999xi}
Bleicher, M.; others.
\newblock {\em J. Phys.} {\bf 1999}, {\em G25},~1859--1896.

\bibitem[Bhalerao \em{et~al.}(2015)Bhalerao, Ollitrault, Pal, and
  Teaney]{Bhalerao:2014mua}
Bhalerao, R.S.; Ollitrault, J.Y.; Pal, S.; Teaney, D.
\newblock {\em Phys. Rev. Lett.} {\bf 2015}, {\em 114},~152301.

\bibitem[Aamodt \em{et~al.}(2011)Aamodt et~al.]{Aamodt:2010cz}
Aamodt, K.; others.
\newblock {\em Phys. Rev. Lett.} {\bf 2011}, {\em 106},~032301.

\bibitem[Adam \em{et~al.}(2016)Adam et~al.]{Adam:2015ptt}
Adam, J.; others.
\newblock {\em Phys. Rev. Lett.} {\bf 2016}, {\em 116},~222302.

\bibitem[Sirunyan \em{et~al.}(2017)Sirunyan et~al.]{Sirunyan:2017gyb}
Sirunyan, A.M.; others.
\newblock {\em Phys. Rev.} {\bf 2017}, {\em C96},~064902.

\bibitem[Schenke \em{et~al.}(2010)Schenke, Jeon, and Gale]{Schenke:2010nt}
Schenke, B.; Jeon, S.; Gale, C.
\newblock {\em Phys. Rev.} {\bf 2010}, {\em C82},~014903.

\bibitem[Schenke \em{et~al.}(2012)Schenke, Jeon, and Gale]{Schenke:2011bn}
Schenke, B.; Jeon, S.; Gale, C.
\newblock {\em Phys. Rev.} {\bf 2012}, {\em C85},~024901.

\end{thebibliography}



\end{document}